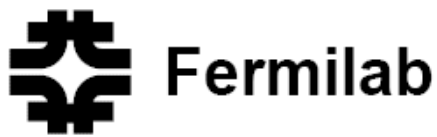

# ACCELERATOR SYSTEM DESIGN, INJECTION, EXTRACTION AND BEAM-MATERIAL INTERACTION: WORKING GROUP C SUMMARY REPORT[*†]

N.V. Mokhov[#], Fermilab, Batavia, IL 60510, U.S.A.
D. Li, LBNL, Berkeley, CA 94720, USA

## Abstract

The performance of high beam power accelerators is strongly dependent on appropriate injection, acceleration and extraction system designs as well as on the way interactions of the beam with machine components are handled. The experience of the previous ICFA High-Brightness Beam workshops has proven that it is quite beneficial to combine analyses and discussion of these issues in one group. A broad range of topics was presented and discussed at the Working Group C sessions at the HB2012 Workshop. Highlights from the talks, outstanding issues along with plans and proposals for future work are briefly described in this report.

___
[*]Work supported by Fermi Research Alliance, LLC under contract No. DE-AC02-07CH11359 with the U.S. Department of Energy.
[†]Presented paper at the 52[nd] ICFA Advanced Beam Dynamics Workshop on High-Intensity and High-Brightness Hadron Beams, HB2012, Beijing, September 17-21, 2012.
[#]mokhov@fnal.gov

# ACCELERATOR SYSTEM DESIGN, INJECTION, EXTRACTION AND BEAM-MATERIAL INTERACTION: WORKING GROUP C SUMMARY REPORT

D. Li, LBNL, Berkeley, CA 94720, USA,
N.V. Mokhov, FNAL, Batavia, IL 60510, USA

## INTRODUCTION

The performance of high beam power accelerators is strongly dependent on appropriate injection, acceleration and extraction system designs as well as on the way interactions of the beam with machine components are handled. The experience of the previous ICFA High-Brightness Beam workshops has proven that it is quite beneficial to combine analyses and discussion of these issues in one group, WG-C at this Workshop. A broad range of topics was presented and discussed in twenty talks at four WG-C sessions as well as at two joint WG-A/C and WG-B/C sessions. The presentations are listed at the end of this report. Highlights from these talks, outstanding issues along with plans and proposals for future work are briefly described in the sections below.

## INJECTION

Stripping foils – carbon or aluminum oxide - is the standard techniques for the H$^-$ injection in the existing machines and projects under consideration. As an example, Fig. 1 shows a typical layout of the system. At ISIS, the serially powered dipoles are used to generate a 45mrad – 65mm vertical bump. Horizontal painting on a 50-μg/cm$^2$ thick Al$_2$O$_3$ foil is done via the closed orbit movement.

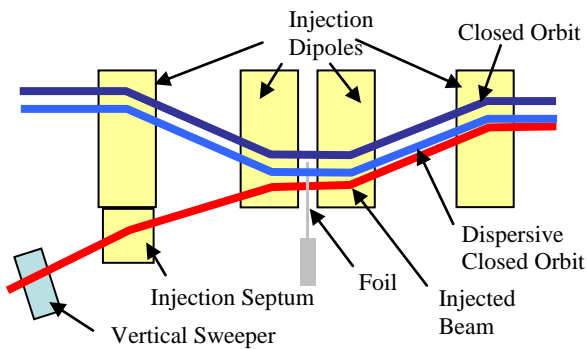

Fig. 1: Schematic layout of ISIS stripping foil injection.

Impressive analysis is performed at SNS to reveal the role of multiple and single Coulomb scattering, energy loss, elastic and inelastic nuclear interactions in stripping foils as a source of beam loss in the machine. The space charge effect is one of the phenomena specific to the high-power accelerators. Detailed 2D/3D ORBIT simulations done for ISIS and Project X allowed quantification of this effect for the technique performance, prediction of beam loss distributions and foil heating. A trick proposed by the ISIS–JPARC collaboration - use of a double-layer foil set – results in a 20% reduction of the peak temperature compared to a standard single-foil setup.

Attention to details and several modifications of the injection system within the Proton Improvement Plan (PIP) would double the 8-GeV proton production in the Fermilab Booster. PIP includes: aperture improvement, better orbit control, magnet re-alignment, a notcher relocation from the L5 straight section to the L12 one along with changing its action from vertical to horizontal, implementation of new stronger correctors for magnetic cogging, switch to 2-stage collimation as was designed originally, and an improved radiation protection scheme.

## BEAM LOSS, COLLIMATION AND EXTRACTION

Comprehensive studies of beam loss and collimation in the ESS Linac have demonstrated difficulties in this area for linear accelerators. The TraceWin tracking simulations were performed to propagate quadrupole and cavity errors allowing optimization of the scheme and loss limit on a graphite collimator. The findings in the MEBT studies are: halo growth occurs in the last half of the MEBT (sometimes in the final 10-20 cm); the standard scheme of two collimators separated by 90 deg is not the optimum for the ESS MEBT; the phase advance of an individual particle (angle in the normalized phase space) depends on its initial position due to a strong space charge; the angular distribution of halo particles is not uniform.

A drastic underestimation of equipment activation due to beam loss in the ESS RFQ (<3 MeV) and DTL (<79 MeV) was found compared to that predicted by the "1 W/m" rule. NVM, as a co-author of that rule, pointed out that "The **1 W/m rule** for beam loss doesn't apply here as we derived it for continuous loss of $E_p$ > 100-200 MeV beam resulting in contact dose (30 day/1 day) of 0.5-1 mSv/hr on an outer surface of a typical (massive) accelerator magnet" (Beam Halo and Scraping, Ed. N.V. Mokhov, W. Chou, 7$^{th}$ ICFA Workshop on High Intensity High Brightness Hadron Beams, Lake Como, Wisconsin, 13-15 Sep. 1999).

The amazing performance of the most powerful PSI accelerator complex is based – among other things – on excellence and innovations in the collimation, extraction and target systems. The power of the 590-MeV proton beam is 1.3 MW, with the peak recently raised to 1.4 MW. The electrostatic extraction channel (0.05-mm tungsten stripes, 16-mm gap, 920-mm length, 8.2-mrad deflection) provides the efficiency of 99.98%. A radiation-cooled graphite target wheel (450-mm diameter, 40-mm thick, 1 revolution per second) performs very well under extreme conditions (20 kW/mA power deposition, 1700K operation temperature, 0.1 DPA/Ah damage rate). The design was further improved by implementing the diagonal gaps to allow dimensional changes under irradiation. The collimation system also performs well. The KHE2 copper collimator gets 150 kW of the beam power and is in operation for more than 20 years. The maximum operation temperature is 653K for a 2-mA beam, the measured residual dose near KHE2 is 500 Sv/h. The beam-induced radiation damage of the collimator material observed after 20 years of operation (120Ah total beam charge) is shown in Fig. 2.

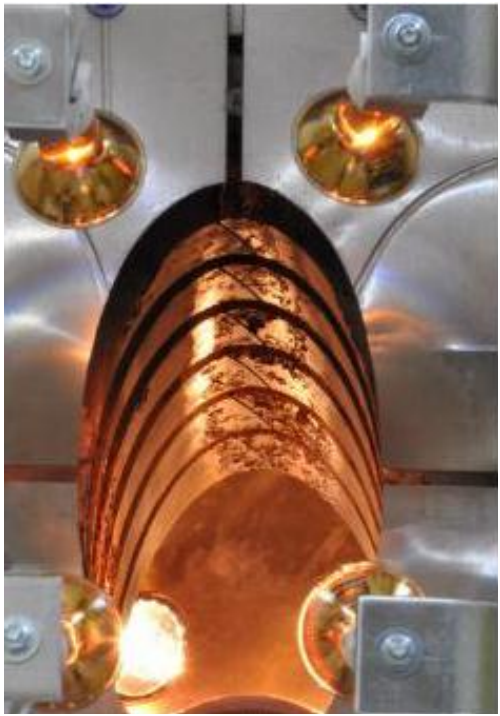

Fig. 2: PSI KHE2 collimator after 20 years of operation.

A detailed study is performed on halo collimation of the intermediate charge-state heavy ions. The earlier idea of using stripping foils in a combination with beam optical elements was further refined at GSI. The stripping foil for halo collimation is placed in the slow extraction area in SIS-100 with stripped ions being intercepted by the two warm quadrupoles.

## NEW CHALLENGING PROJECTS

A high-performance *Mu2e experiment* has been proposed at Fermilab to search for the muon to electron conversion in the field of a nucleus. A schematic layout of the experiment is shown in Fig. 3. Muons are created from decaying pions generated by a proton beam striking a production target. The muons are transported to an aluminum target in the detector solenoid where they are captured and decay. The Mu2e detector is quite challenging, where statistical filter and detector strategy are needed. The detection window is about 925 ns, right after backgrounds have died away. Nevertheless, the extinction sensitivity must be at $10^{-10}$ levels, which are four orders of magnitude better than the current limit. Operation of the experiment is scheduled to start in FY2020.

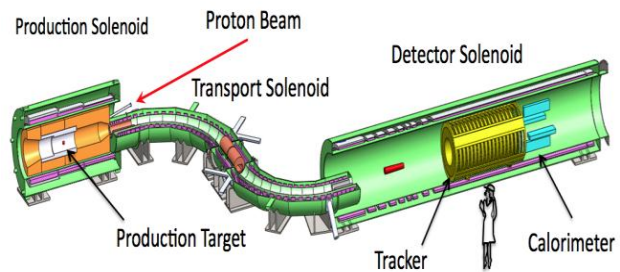

Fig. 3: Schematic layout of the Mu2e experiment at Fermilab.

*Michigan State University Rare Isotope Re-Accelerator* is under construction for the FRIB (Facility for Rare Isotope Beams) project at MSU (Michigan State University), as shown in Fig. 4. It will provide unique low-energy rare isotope beams as a test bed for the FRIB accelerator technology, which includes EBIT (Electron-Beam Ion Trap), RFQ, SRF cavities and their cryomodules and beam diagnostics. The accelerator complex is currently under commissioning, and will be able to accelerate rare isotope beams to ~ 3 MeV/u in 2013 to meet the strong demands for nuclear physics and astrophysics research programs.

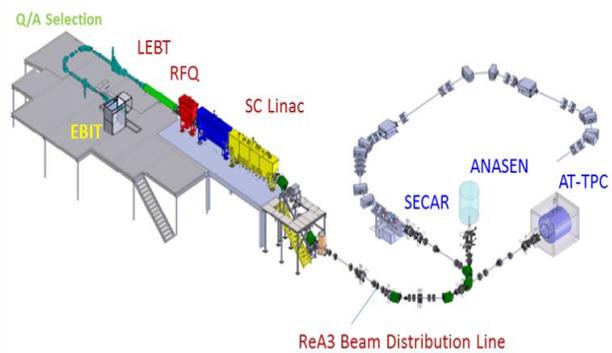

Fig. 4: Layout of the MSU Rare Isotope Re-Accelerator.

*Project X* is a proposed accelerator complex at Fermilab that is capable of producing multi-MW proton beams in a few GeV energy range for future high-energy physics programs. Fig. 5 shows the latest layout proposed for Project X. The accelerator complex consists of an H- DC ion source, a 162.5 MHz normal conducting 2.1 MeV CW RFQ accelerator, low-beta superconducting HWR (Half Wavelength Resonator) cavities and a 3 GeV superconducting CW lianc at 650 MHz.

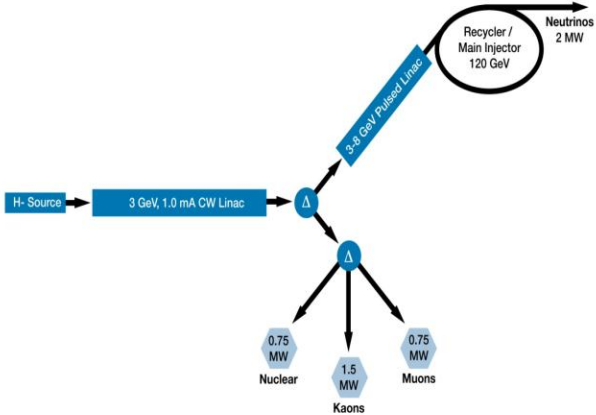

Fig. 5: A schematic layout of the accelerator complex of the proposed Project X at Fermilab, with proposed experiments and upgrade paths.

A staged approach scheme with a 1-GeV injection design has been proposed. Beam loss control for the Fermilab Main Injector has also been studied and presented. In collaboration with Fermilab, LBNL (Lawrence Berkeley National Laboratory) is responsible for the R&D and design studies of the front-end system for Project X, which includes an H- ion source, LEBT and the CW RFQ accelerator. Time dependent neutralization and beam dynamics due to chopper in LEBT is critical, numerical simulation studies using WARP 3D code have been conducted at LBNL. Preliminary benchmark experiments started and will continue to validate the simulation model.

Significant investment and progress have been made on superconducting RF technology at Fermilab over the last decade primarily for ILC at 1.3 GHz. Project X requires low-beta CW SRF cavities at lower frequencies, 325 MHz and 650 MHz for beta from 0.11 to 0.9, respectively. The SRF linac technologies developed for ILC such as SRF material R&D, cavity fabrication, processing, testing and cryostat can be used for Project X. Recent progresses include 325-MHz prototype Spoke cavities and 650 MHz low-medium beta elliptical cavities. A world record cavity $Q_0 \sim 7.5 \times 10^{10}$ was achieved recently in a testing for a 1.3 GHz cavity with NbN coating, and traditional Q-slope curve was not observed. More R&D programs and testing have been planned to study and understand this discovery. A higher $Q_0$ is desirable for any CW operation machines; it has huge impact on the cost of required RF and cryogenic power. In addition to the SRF technology development for Project X at Fermilab, a complete overview of the state-of-art SRF technology, challenges and future development directions have been summarized.

## NEW TECHNIQUES

Several high-potential techniques have been presented and discussed at the WG-C sessions: (1) *A combination of electron and stochastic cooling* at the same high beam energy; the key problems of a 2-MeV electron cooler were experimentally verified in Novosibirsk; the cooler is ready for assembly and commissioning at COSY. (2) *Crystal collimation:* after the Tevatron shutdown, most of activities in this field were moved to CERN; impressive results of the beam tests at SPS and plans for tests in the LHC were presented. (3) *Hollow electron beam collimation:* magnetically confined hollow electron beams are a safe and flexible technique for halo control in high-power and high-stored energy accelerators; it is a material-less soft complement to two-stage collimation; very promising for LHC; Tevatron experiments have provided an experimental foundation.

## MATERIALS

New study confirmed that energetic proton fluence thresholds are a reality for carbon-based lattices. There is a significant variability between the graphite grades in the way that graphite responds to irradiation. The non-destructive testing has shown a great potential in assessing the damage annealing. The fast neutron exposure to fluences similar to those for protons has just been completed and will provide a good correlation between the different irradiating species

It was shown that the ion-induced disordering of graphite is different from that caused by neutrons with respect to swelling, stress concentrators, bending, hardening, degradation of thermal conductivity and fatigue resistance. A steep degradation of properties takes place at doses corresponding to the ion track overlapping that is given by ion track size and depends on ion mass and energy. The high temperature (above 1000°C) operation of graphite extends its lifetime due to defect recovery. Fatigue induced by cyclic thermo-mechanical loading reduces the lifetime. This is illustrated in Fig. 6 for the damage observed in graphite films at GSI. A 4.8 MeV/u $^{197}$Au beam was used to get an accumulated fluence of $10^{14}$ cm$^{-2}$. The first test was quasi-continuous irradiation at 48 Hz while the second one was pulsed irradiation with a repetition rate of 0.4 Hz. The difference is quite striking.

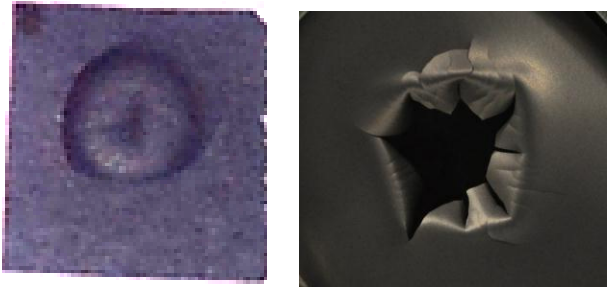

Fig. 6: Radiation damage caused in graphite films by ion irradiation to the same fluence in the quasi-continuous (left) and pulsed (right) modes.

## UNCERTAINTIES IN SIMULATIONS

Nowadays all the aspects of beam interactions with accelerator system components are addressed in sophisticated Monte-Carlo simulations benchmarked - wherever possible - with dedicated beam tests. Fig. 7 shows the energy and Intensity Frontier applications where the state-of-the-art codes, such as FLUKA and MARS, are used.

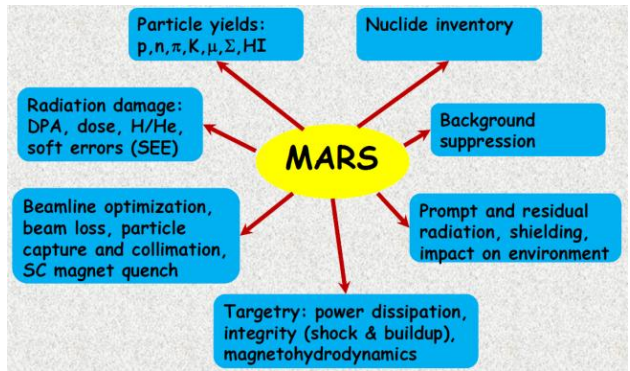

Fig. 6: Major applications at the Intensity Frontier of particle-matter interaction MARS code.

Predictive power, capabilities and reliability of the major particle-matter interaction codes used in accelerator applications are quite high. At the same time, analysis of the status and uncertainties in modeling of radiation effects caused by high-power beams has revealed some issues. The most fundamental one is particle production in nuclear interactions that is the heart of all such simulations and the key for collimator, target and other machine component design as well as fixed target and collider experiment planning. Overall, the situation is quite good for beam energies below 1 GeV and above 10 GeV with accuracy of predictions being at a 20% level in most cases. At intermediate energies - most interesting for the Intensity Frontier – there are substantial theoretical difficulties. Moreover, the experimental data contradict each other at these energies. The main problem is with the low-energy pion production that is crucial, e.g., for all the Project X experiments. Accuracy of beam-induced macroscopic effect predictions today is

- Energy deposition effects (instantaneous and accumulated) < 15%.
- Hydrogen/Helium gas production and DPA: ~20% (with similar DPA models) to a factor of 2; still need a better link of calculated DPA to the observed changes in material properties.
- Beam loss generation and collimation: quite good (Tevatron, J-PARC, LHC).
- Radiological issues (prompt and residual): a factor of 2 for most radiation values if all details of geometry, materials composition and source term are taken into account.

## NEEDS AND ACTION ITEMS

The following needs and action items were identified at the WG-C sessions:

- More work is still needed toward a high-performance stripping injection including code verification and fully 3D simulations.
- Beam tests at LHC to reveal the high potential of the hollow electron beam collimation.
- Full 3D models of the entire machine integrated for beam loss and shower simulations.
- Linking CAD and simulation codes; geometry exchange tools.
- Low-energy pion production. How to resolve the HARP and HARP-CDP disagreement?
- Materials beam tests: cryogenic temperatures, high-energy protons, annealing, and atmosphere.
- Design of "Dream materials" for foils, targets, collimators and beam dumps; nano-structures.
- Moving from the calculated dose and DPA to changes in material properties: ready for coupling shower simulation codes (FLUKA, MARS) and "materials" modeling codes.

## WG-C PRESENTATIONS

1. S. Montesano "Status and results of the UA9 crystal collimation experiment at the CERN SPS"
2. D. Johnson "Injection design for Fermilab Project X"
3. Y. Yuan "Study of intense beam injection and extraction of heavy ion synchrotron"
4. B. Brown "Beam loss control for the Fermilab Main Injector"

5. X. Wu "The design and commissioning of the accelerator system of the Rare Isotope Reaccelerator – ReA3 at Michigan State University"
6. V. Reva "High-energy electron cooling"
7. R. Miyamoto "Beam loss and collimation in the ESS linac"
8. D. Reggiani "Extraction, transport and collimation of the PSI 1.3 MW proton beam"
9. F. Garcia "Current and planned high proton flux operations at the FNAL Booster"
10. B. Pine "Injection and stripping foil studies for a 180-MeV injection upgrade at ISIS"
11. I. Strasik "Collimation of ion beams"
12. G. Stancari "Beam halo dynamics and control with hollow electron beams"
13. N. Simos "LBNE target material radiation damage studies using energetic protons of the BLIP facility"
14. N. Mokhov "Radiation effect modeling at intensity frontier: status and uncertainties"
15. M. Tomut "Understanding ion induced radiation damage in target materials"
16. E. Prebys "Proton beam inter-bunch extinction and extinction monitoring for the Mu2e experiment"
17. A. Facco "SRF technology challenge and developments"
18. R. Kephart "SRF cavity research for Project X"
19. Q. Li "Beam dynamics studies of H- beam chopping in a LEBT for Project X"
20. L. Sun "Intense high charge state heavy ion beam production for the advanced accelerators"